\documentstyle[prd,aps,preprint,floats,epsfig]{revtex}
\tighten

\begin{document}
\draft
 
\pagestyle{empty}

\preprint{
%\noindent
%\begin{minipage}[t]{3in}
%\begin{flushleft}
\today \\
%\end{flushleft}
%\end{minipage}
%\hfill
\begin{minipage}[t]{3in}
\begin{flushright}
%LBNL--aaaaa \\
%UCB--PTH--02/yy \\
%hep-ph/0205xxx \\
month May 2002
\end{flushright}
\end{minipage}
}

\title{Helicity conservation in inclusive nonleptonic decay 
$B\to VX$:\\ Test of long-distance final-state interaction}

\author{
Mahiko Suzuki
%\thanks{Work supported in part by the Director, Office of Energy
%Research, Office of High Energy and Nuclear Physics, Division of High
%Energy Physics of the U.S. Department of Energy under Contract
%DE--AC03--76SF00098 and in part by the National Science Foundation under
%grant PHY--0098840.}
}
\address{
Department of Physics and Lawrence Berkeley National Laboratory\\
University of California, Berkeley, California 94720
}

%\thanks{Work supported by the Department of Energy under Contract
%DE--AC03--76SF00515.}

\date{\today}
\maketitle

\begin{abstract}

   The polarization measurement in the inclusive $B$ decay 
provides us with a simple test of how much the long-distance 
final-state interaction takes place as the energy of the 
observed meson varies in the final state. We give the 
expectation of perturbative QCD for the energy dependence
of the helicity fractions in a semiquantitative form. Experiment 
will tell us for which decay processes the perturbative QCD
calculation should be applicable.

\end{abstract}
%\pacs{}
%\newpage
\pagestyle{plain}
\narrowtext
                                                                   
\setcounter{footnote}{0}

\section{Introduction}

      It is of crucial importance to know how much 
long-distance final-state interaction (LDFSI) occurs in 
$B$ decay. If LDFSI plays a significant role, we have no 
first-principle method to compute decay amplitudes. Arguments 
have been presented in favor of short-distance (SD) dominance 
for the two-body decay in which a fast quark-antiquark pair 
moves almost collinearly in a colorless lump. Based on this color 
screening picture\cite{BJ}, the perturbative QCD computation 
has been carried out for the two-body $B$ decay\cite{Beneke,Li}. 
Even if the SD dominance argument is valid in the infinite $B$
mass limit, a quantitative question exists about the accuracy 
of the perturbative QCD calculation since the $B$ meson mass 
is only 5.3 GeV in the real world.  When the final mesons are 
highly excited states, the velocities of the mesons are less 
fast and the quarks inside them have larger transverse momenta. 
We expect that the SD dominance is accordingly less accurate 
in such decays. In the large limit of the excited meson mass, 
the LDFSI should play a major role in determining the final 
state. We would like to verify experimentally the SD dominance 
in the two-body decay and see how the SD dominance disappears 
as the meson slows down in the inclusive decay.

One of the cleanest ways to test breakdown of the SD dominance 
or presence of LDFSI directly with experiment is to measure the 
helicity of a fast flying meson in the final state\cite{Suzuki}. 
Since SD interactions do no flip helicities of light quarks 
($u,d,s$), a fast light meson carries a memory of the quark 
helicities if no LDFSI enters. Because of the specific form of 
the weak interaction in the Standard Model, a fast light meson 
with spin must be polarized in the zero helicity state up 
to $O(1/M_B^2)$ in probability, when other hadrons fly away 
approximately to the opposite directions. One can determine the 
$h=0$ fraction of the meson by measuring the angular distribution 
of the its decay products. In fact, this selection rule is so 
robust that it would be valid even if the right-handed $W$-boson
contributes to weak decays. It breaks down most likely by LDFSI,
if at all. 

  Imagine that such polarization measurement is made for the 
inclusive decay $B\to\rho X$ in which $X$ is a highly excited 
meson state ($\overline{q}q$) or a multi-quark hadronic state. 
As the invariant mass $m_X$ increases, it becomes more likely
that LDFSI takes place between $\rho$ and $X$. If so, we shall 
start seeing production of the $\rho$ meson in the $h=\pm 1$ 
states. By measuring the $\rho$ helicity as a function of 
$m_X^2$ or equivalently as a function of the $\rho$ energy 
in the $B$ rest frame, we can determine from experiment how 
much LDFSI enters the decay as $\rho$ slows down or how much 
the color screening breaks down.

For the two-body decay, the polarization measurement is 
possible only when both final mesons have nonzero spins, 
for instance, $B\to 1^-1^-$. Meanwhile, most decay modes that 
are easily identifiable and high in branching fraction are 
$B\to 0^-0^-$ and $1^-0^-$. Nonetheless, the polarization test 
will have a direct impact on these dominant decay modes of 
the $B$ meson in the following way. In the charmless $B$ decay,  
the two-body decays $B\to \pi\pi$ and $\rho\pi$ are among 
the decay modes of primary interest from the viewpoint of 
CP violation. If our proposed test reveals that the $h=0$ 
state dominates in $B\to\rho\rho$, $\rho\omega$, and so forth, 
we shall feel more confident with computing the tree and penguin 
amplitudes of $B\to\pi\pi$, $\rho\pi$ in perturbative QCD. 
If on the contrary the $h=0$ dominance is substantially 
violated in $B\to\rho\rho$, $\rho\omega$, we should not trust 
the perturbative method of calculation for $B\to\pi\pi$, 
$\rho\pi$. In this case the only recourse would be to 
determine the $B\to \pi\pi$ amplitudes by experiment 
alone\cite{Gronau} without help of theoretical computation. 
And little could be done for $B\to\rho\pi$ with isospin 
invariance alone.  The test proposed here is not for 
inventing a new method of calculation of decay amplitudes, 
but for learning from experiment for which decay
modes we may perform the perturbative QCD calculation.

\section{Kinematics of $B\to V X$}

   We consider the inclusive $B$ decay into a vector meson $V$
of $J^P=1^-$;
\begin{eqnarray} 
   B({\bf P})  \to  &V({\bf q}, h)&  + X(p_X), \nonumber \\
            &\searrow & a({\bf k}_1) + b({\bf k}_2) + X(p_X),
\end{eqnarray}
where $a$ and $b$ are spinless decay products of $V$ ($m_b\neq m_b$ 
in general). Here we have $B\to \rho X$, $K^* X$, and 
$\phi X$ in mind. The inclusive decay rate is written 
in the covariant form as
\begin{eqnarray}
  4(2\pi)^6k_{10}k_{20}\frac{d\Gamma}{d^3k_1d^3k_2}&=&\nonumber\\
   \sum_{ij}\int\frac{d^3{\bf q}}{4(2\pi)^3 q_0P_0}
   \frac{g_{ab}^2}{2m_V\Gamma_V} &(2\pi)^4&\delta^4(k_1+k_2-q)
   (\epsilon_i\cdot k_1-k_2)(\epsilon_j^*\cdot k_1-k_2)
     \epsilon^{\mu*}_i T_{\mu\nu}\epsilon^{\nu}_j, \label{diff}
\end{eqnarray}
where $\Gamma_V$ is the decay width of $V$, $g_{ab}$ is
the decay coupling constant of $V$ defined by 
$L_{int}=ig_{ab}(\phi_a^*\stackrel{\leftrightarrow}{
\partial_{\mu}}\phi_b^*) V^{\mu}$, and the subscript of the 
polarization vector $\epsilon$ refers to three helicity states
of $V$. The covariant tensor $T_{\mu\nu}$ is the inclusive 
structure function defined by
\begin{equation}
   T_{\mu\nu}(m_X^2) = 4q_0P_0 \sum_X (2\pi)^4\delta^4(q+p_X-P)
     \langle B({\bf P})|H_{int}|V({\bf q},j)X\rangle
 \langle V({\bf q},i)X|H_{int}|B({\bf P})\rangle.\label{T}
\end{equation}
where the states are normalized as $\langle{\bf p}|{\bf p}'\rangle=
(2\pi)^3\delta({\bf p}-{\bf p}')$ without $2E_{{\bf p}}$.
The general tensor form of $T_{\mu\nu}$ is
\begin{equation}
    T_{\mu\nu} =  - g_{\mu\nu}A(m_X^2) 
         + \frac{1}{M_B^2}P_{\mu}P_{\nu}B(m_X^2)
    + \frac{i}{M_Bm_V}\varepsilon_{\mu\nu\kappa\lambda}
    P^{\kappa}q^{\lambda} C(m_X^2),                \label{T2}
\end{equation}
where $m_X^2=(P-q)^2$ and the antisymmetric unit tensor is defined
as $\varepsilon_{0123}=-1$. The scalar structure functions $A\sim C$
are the absorptive parts of the analytic functions of variable 
$m_X^2$ that are regular except on the segments of the real axis 
in the complex $m_X^2$ plane if $V$ is treated as (approximately) 
stable. In particular, $A\sim C$ are nonsingular ($\neq\infty$)
in the physical region of the decay.  

The helicity amplitudes $H_h$ for $B\to V_h X$ in 
the $B$ rest frame can be expressed in terms of $A\sim C$ as
\begin{eqnarray}
    H_0 &=& A +\frac{{\bf q}^2}{m_V^2}B, \nonumber \\
    H_{\pm 1} &=& A \mp \frac{|{\bf q}|}{m_V}C. \label{helicity}
\end{eqnarray} 
In contracting $T_{\mu\nu}$ with $\epsilon$, we must not make the 
approximation $\epsilon^{\mu}({\bf q})\simeq q^{\mu}/m_V$ as 
we often do in the exclusive two-body decay $B\to V_1V_2$ where 
$g_{\mu\nu}\epsilon_1^{\mu}\cdot\epsilon_2^{\nu} \simeq 
(q_1\cdot q_2)/m^2$. Because $g_{\mu\nu}\epsilon^{\mu^*}
\epsilon^{\nu}= -1$ while $g_{\mu\nu}q^{\mu}q^{\nu}/m_V^2= +1$ 
in the inclusive decay kinematics.

Carrying out the summation over the helicities in Eq.(\ref{diff}) 
with Eq.(\ref{T2}), we obtain the differential decay rate with 
respect to the direction of ${\bf k}_1$ and the energy of $V$. 
The result is:
\begin{equation}
         \frac{d\Gamma(B\to VX \to abX)}{
             dq_0 d\cos\theta}\Bigg|_{B\;{\rm at}\;{\rm rest}} =
 \frac{g_{ab}^2|{\bf q}||{\bf k}_{cm}|^3}{32\pi^3m_V^2\Gamma_V}
 \biggl[A(m_X^2)+\frac{{\bf P}^2}{M_B^2}B(m_X^2)\cos^2\theta\biggr], 
                              \label{angular}
\end{equation}
where $q_0$ is the energy of $V$ in the rest frame of $B$, 
which is related to $m_X$ by $m_X^2 = M_B^2+m_V^2-2M_Bq_0$ 
so that $d\Gamma/dq_0 = 2M_Bd\Gamma/dm_X^2$, ${\bf k}_{cm}$ 
is the momentum of $a$ in the rest frame of $V$, ${\bf P}$ 
is the momentum of $B$ measured in the rest frame of $V$, and 
$\theta$ is the angle of ${\bf k}_{cm}$ measured from the 
direction of ${\bf P}$, namely, $({\bf P}\cdot{\bf k}_{cm})
=|{\bf P}||{\bf k}_{cm}|\cos\theta$.
 
We make two remarks on Eq.(\ref{angular}). Since the decay 
products $a$ and $b$ are spinless, the structure function of 
the $V\to ab$ decay, $g_{ab}^2(k_1-k_2)^{\mu}(k_1-k_2)^{\nu}$, 
is symmetric under $\mu\leftrightarrow\nu$ so that the function 
$C(m_X^2)$ does not enter the differential decay rate. It means 
according to Eq.(\ref{helicity}) that we cannot separate the 
$h=-1$ decay from the $h=+1$ decay in this process. In order 
to distinguish between $h=\pm 1$, we would have to choose a decay
in which $J\neq 0$ for $a$ or $b$ and to measure the helicity 
of $a$ or $b$ through its decay. For instance, the triple product 
${\bf q}\cdot({\bf k}_1\times{\bf k}'_1)$ in the sequence of decays 
$B\to a_2({\bf q}) X\to\pi({\bf k}_1)\rho({\bf k}_2)X\to
\pi({\bf k}_1)\pi({\bf k}'_1)\pi({\bf k}'_2)X$ contains such an 
information.\footnote{
Such measurement was actually proposed to determine the 
photon helicity in $B\to\gamma K_1\to\gamma K\pi\pi$\cite{triple}.
The strong phases due to the overlapping resonances are needed to 
detect the triple product.}
The other comment is on the slow limit of $V$. In the limit of 
${\bf q}\to 0$ in Eq.(\ref{helicity}), distinction among three 
different helicity states of $V$ disappears for an obvious reason 
and all helicity functions $H_h$ ($h=1, 0,-1$) are given by 
$A(m_X^2)$ since $B(m_X^2)$ and $C(m_X^2)$ stay finite there:
\begin{equation}
   H_1 + H_{-1} \to 2H_0, \;\; H_1 - H_{-1}\to 0\;\; {\rm as}
                \; {\bf q}\to 0.   \label{limit}
\end{equation} 
In this limit only the $A(m_X^2)$ function survives in the
differential decay rate of Eq.(\ref{angular}), as we expect, 
since ${\bf q}\to 0$ means ${\bf P}\to 0$.

Finally, let us express the differential decay rate in terms
of $H_h$, noting that $|{\bf P}|/M_B=|{\bf q}|/m_V$ by the 
transformation between the $B$ rest frame and the $V$ rest frame. 
The result is
\begin{equation}
         \frac{d\Gamma(B\to VX\to abX)}{
             dq_0 d\cos\theta}\Bigg|_{B\;{\rm at}\;{\rm rest}} =
  \frac{g_{ab}^2|{\bf q}||{\bf k}_{cm}|^3}{32\pi^3m_V^2\Gamma_V}
 \biggl[H_0\cos^2\theta+\frac{1}{2}(H_1+H_{-1})\sin^2\theta\biggr].
                                \label{d}
\end{equation} 
We are able to separate between the longitudinal ($h=0$) and 
transverse ($h=\pm 1$) polarization decay with the angular 
distribution of Eq.(\ref{d}). Experiment will show us how the 
$h=0$ dominance goes away as $m_X$ increases in the inclusive 
decay $B\to VX$. If the transverse polarization appears beyond 
the corrections to be discussed in the subsequent sections, 
it will be a clear evidence for LDFSI.

\section{Longitudinal polarization dominance} 

       For the weak interaction of the Standard Model, the 
zero-helicity function $H_0$ should dominate over all other 
$H_{\lambda}$ for small $m_X$, if the strong interaction 
corrections are entirely of short distances except at hadron 
formation. We explain this rule for two-body decays\cite{Suzuki}, 
discuss the mass and orbital motion corrections to the rule, 
and extend it to the inclusive decay $B\to VX$. Our argument 
is based on the standard assumptions made in the perturbative 
calculation including the light-cone formulation of mesons in 
$\overline{q}q$. The helicity selection rule should break down 
for sufficiently large values of $m_X$. At which value the rule 
starts showing a significant departure from the $h=0$ dominance 
will provide us with a quantitative measure of accuracy of the
perturbative QCD calculation. We first discuss the charmless 
decay and then move on to the decays with charm. 

\subsection{Meson helicity and helicities of $\overline{q}q$}

In the nonleptonic $B$ decay a pair of $\overline{q}q$ is 
produced by weak interaction nearly in parallel to form 
an energetic meson. In the case of a vector meson $(^3S_1)$, 
we may approximate the $\overline{q}q$ pair to be literally 
in parallel by ignoring a tiny $^3D_1$ component. For excited 
mesons such as $J^P=2^+(^3P_2)$, the transverse motion of $q$ 
and $\overline{q}$ must be taken into account. It gives rise to 
an orbital angular momentum $l$ between $q$ and $\overline{q}$
as well as to the meson mass. This angular momentum is
part of the meson spin. By simple kinematics, however, 
the state of $l_z=0$ dominates over all others when a meson 
moves fast. That is, to the lowest order we may leave out the 
orbital motion of $\overline{q}q$ inside a meson even for an 
excited meson state with $l\neq 0$. Let us make this statement
quantitative.

In the classical picture, the orbital angular momentum 
vector is squashed to the plane perpendicular to the meson 
momentum when a meson moves fast. To see it in quantum theory, 
let us expand the plane wave $e^{i{\bf p}\cdot{\bf r}}$ of a 
quark in the spherical harmonics for ${\bf p}$ off the 
direction of the meson momentum ${\bf q}=|{\bf q}|\hat{z}$. 
Defining the directions of the vectors as
\begin{eqnarray}
 {\bf r} &=& r(\sin\vartheta\cos\varphi,\; \sin\vartheta\sin\varphi,
               \; \cos\vartheta), \nonumber \\
 {\bf p} &=& |{\bf p}|
       (\sin\vartheta'\cos\varphi',\; \sin\vartheta'\sin\varphi',\;
               \cos\vartheta'), \nonumber \\
       \hat{{\bf r}}\cdot\hat{{\bf p}} &\equiv& \cos\gamma.
\end{eqnarray}
We obtain by use of the well-known formulae the expansion of 
the plane wave in the form
\begin{eqnarray}
  e^{i{\bf p}\cdot{\bf r}} 
 &=& \sum_l (2l+1)i^l j_l(|{\bf p}|r)P_l(\cos\gamma), \nonumber \\
 &=& 4\pi\sum_l i^l j_l(|{\bf p}|r) \sum_{m=-l}^l 
 Y_{lm}^*(\vartheta',\varphi')Y_{lm}(\vartheta,\varphi).\label{wave}
\end{eqnarray}
Treating $\vartheta'\simeq |{\bf p}_T|/|{\bf p}|$ as small, 
we expand $Y_{lm}^*(\vartheta',\varphi')$ around $\vartheta'=0$.
Then Eq.(\ref{wave}) turns into 
\begin{equation}
  e^{i{\bf p}\cdot{\bf r}} \simeq
   \sum_l\sqrt{4\pi(2l+1)}i^lj_l(|{\bf p}|r)
     \sum_{m=-l}^l\frac{(-1)^{m+|m|}}{2^{|m|} |m|!}
 \sqrt{\frac{(l+|m|)!}{(l-|m|)!}} e^{-im\varphi'}\vartheta'^{|m|} 
          Y_{lm}(\vartheta, \varphi).
\end{equation}
In the sum over $l_z$ (denoted by $m$ above), the amplitudes 
of $l_z\neq 0$ are suppressed by $\vartheta'^{|l_z|}\simeq 
(\frac{1}{2}m_T/E)^{|l_z|}$ where $m_T$ stands for the 
transverse meson mass ($\simeq\sqrt{\frac{2}{3}}\times$ meson 
mass). Repeat the argument for $\overline{q}$. Projecting
the $\overline{q}q$ state with the quark distribution function 
of meson, we find that the meson helicity consists entirely 
of the quark helicity $h_q+h_{\overline{q}}$ in the fast limit. 
The contribution of the $l_z\neq 0$ states generates a correction 
of $O((\frac{1}{2}m_T/E)^{|l_z|})$ in amplitude for an excited 
meson and a multi-meson state. 

\subsection{Helicity selection rule; charmless decay}

  The fundamental weak interaction is dressed or improved into 
the effective decay operators by the renormalization group 
down to the scale $m_b$. In the Standard Model, the chiral 
structure of the decay operators relevant to the charmless 
decay are $(\overline{b_L}q_L)(\overline{q_L}q_L)+h.c.$ and
$(\overline{b_L}q_L)(\overline{q_R}q_R)+h.c.$, where $q$ stands 
for a light quark. The short-distance interaction below the 
scale $m_b$ does not generate any new chiral structure. It 
can add $\overline{q_L}q_L + \overline{q_R}q_R$ through
quark pair emission by a hard gluon. The chirality of the
spectator quark is indefinite so that it can be either in 
helicity $+\frac{1}{2}$ or $-\frac{1}{2}$ when it forms 
a meson. 

  Let us start with the two-body charmless decay $B\to VM$
($J\geq 1$ for M too). When one of $V$ and $M$ is formed with 
$\overline{q_L}q_L$ or with $\overline{q_R}q_R$, this meson 
is in the $h=0$ state. The angular momentum conservation 
along the decay momenta in the $B$ rest frame requires that 
the helicity of the other meson must also be zero (Fig.1a). 
Therefore $H_0$ dominates in this case. Alternatively with
$(\overline{b_L}q_L)(\overline{q_R}q_R)$, if $\overline{q_L}
q_R (h=+1)$ is combined to form one meson, the other meson 
must be made of the spectator $q_{spec}$ and $\overline{q_R}
(h=-\frac{1}{2})$. Then the net helicity of the second meson 
can be only $0$ or $-1$, which does not match the helicity 
$h=+1$ of the first meson (Fig.1b). Therefore the only two-meson 
state compatible with helicities and the overall angular 
momentum conservation is $V_{h=0}M_{h=0}$. This argument is 
valid only in the limit that the massless $q$ and $\overline{q}$ 
move strictly in parallel and there is no relative motion 
between them inside the meson.

%\input psfig
%\noindent
\begin{figure}[ht]
\hskip 5cm
\epsfig{file=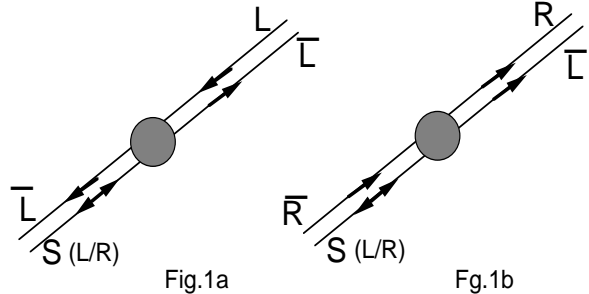,width=8cm,height=4cm} 
\caption{The quark helicities in the two-body $B(\overline{b}q)$
decay. $L$ ($\overline{L}$) and $R$ ($\overline{R}$) denote left
and right chiral quarks (antiquarks), respectively. The spectator
quark $q_{spec}$ is denoted by $S$. The arrows represent
the dominant spin directions. (a) The case of $\overline{b}\to
\overline{q_L}q_L\overline{q_L}$. (b) The case of $\overline{b}
\to\overline{q_L}q_R\overline{q_R}$.
\label{fig:1}}
\end{figure}

\subsection{Mass corrections}

 The relative motion of $q\overline{q}$ generates 
a correction to this helicity selection rule. Since the 
motion of light quarks makes up the entire mass of a 
nonflavored meson, this correction should be $O(|{\bf p}_T|/E) 
= O(\frac{1}{2}m/E)$ in amplitude, where $m$ is 
the meson mass and $E\approx \frac{1}{2}M_B$ for two-body 
decays. When either mass of $V$ and $M$ is large, the 
correction is large and accuracy of the rule is reduced 
accordingly. Let us examine this correction.
  
In the case of the $B(\overline{b}q)$ meson decaying through 
the interaction $(\overline{b_L}q_L)(\overline{q_L}q_L)$, 
the quarks in the final state are
$\overline{q_L}q_L\overline{q_L}q_{spec}$ where $q_{spec}$ 
stands for the spectator. The $h=+1$ state of the meson 
($\overline{q_L}q_L$) can arise from a small opposite 
helicity component of a single $q_L$ while $h=+1$ is 
allowed for the other meson ($\overline{q_L}q_{spec}$) 
thanks to the indefinite helicity of $q_{spec}$. On the
other hand, formation of the $h=-1$ mesons state requires 
the small opposite helicity components of two 
$\overline{q}_L$'s, one in $V$ and one in $M$ (Fig.1a).  
Consequently, $H_1$ arises as the first-order correction 
while $H_{-1}$ can arise only as the second-order 
correction. The same conclusion follows when $B$ decays 
through $(\overline{b_L}q_L)(\overline{q_R}q_R)$ (Fig.1b).
If we define the longitudinal and transverse 
fractions of helicity decay rates by
\begin{equation}
     \Gamma_L = \frac{H_0}{H_1 + H_0 + H_{-1}}, \;\; 
     \Gamma_T = 1 - \Gamma_L,  \label{frac}
\end{equation}
the mass corrections are expressed as $\Gamma_T =
O(m^2/M_B^2)$ and $\Gamma_L=1-O(m^2/M_B^2)$ in the case 
of the two-body decay $B(\overline{b}q)\to 
VM$\cite{Suzuki}.\footnote{
Such a mass correction can be seen in the $U(6)\times U(6)$ 
model calculation of the charmless decay $B\to 1^-1^-$ by
Ali {et al}\cite{Ali}. Cheng {\em et al.} 
recently referred to this correction in their improved 
factorization calculation of $B\to J/\psi K^*$\cite{Cheng}.
Many other model calculations in the past based on
the factorization, however, do not follow this pattern
of mass corrections since vector and axial-vector form
factors were introduced without chiral constraints.} 
Here $m$ is the mass of the meson which does not 
receive the spectator quark or its descendant. 
The reason is obvious from the preceding argument: It is 
the meson formed by the energetic $\overline{q}q$ 
originating from the effective decay interaction that 
primarily determines the helicity state, since the 
helicity of the other side that receives the spectator 
has a twofold uncertainty due to the indefinite spectator 
helicity. The helicity of the meson carrying the spectator 
is constrained by the overall angular momentum conservation. 
In the case of the $\overline{B}(b\overline{q})$ meson, 
the mass corrections to $H_1$ and $H_{-1}$ are interchanged 
in the same argument. 

     We should recall that there is also the $l_z\neq 0$ 
correction of $O(m^2/M_B^2)$ in probability in the case that 
a meson has $l\neq 0$. This correction contributes to $H_1$ 
and $H_{-1}$ in the same order, namely $\vartheta^2$. 
In terms of $\Gamma_T\approx H_1+H_{-1}$ the correction
takes the same form for excited mesons. 

     It is easy to see here that the $h=0$ dominance holds 
even if the right-handed current enters weak interaction.
Only $H_1 > H_{-1}$ or $H_{-1}> H_1$ in the mass correction
depends on $V-A$ or $V+A$. In order to violate the $h=0$ 
dominance, we would need such an exotic weak interaction 
as $b\to q_L\overline{q_R}q_L$. If the $h=0$ dominance
breaks down, therefore, the most likely source is LDFSI.

\subsection{Inclusive charmless decay}
 
The argument in the preceding section can be immediately 
extended to the inclusive decay $B\to VX$ in the case that 
$X$ is described as excited $q\overline{q}$ states. When a 
$q\overline{q}$ pair is created almost collinearly by a  
hard gluon and turns $X$ into a $q\overline{q}q\overline{q}$ 
state, the added pair $\overline{q_L}q_L$ or 
$\overline{q_R}q_R$ has net helicity zero and does 
not contribute to the helicity of $X$ (Fig.2a). 
In this case the previous argument of the $h=0$ dominance is 
unaffected. It can happen alternatively that the hard 
$q$ and $\overline{q}$ are emitted back to back. Imagine, 
for instance, that $q_R$ enters $V$ and $\overline{q_R}$ 
goes into $X$ so that $V\sim \overline{q_L}q_R$ and 
$X\sim\overline{q_R}q_L\overline{q_L}q_{spec}$ (Fig.2b). 
Then the net helicities are $h=+1$ for $V$ and $h=0,-1$ 
for $X$, so the additional hard pair of 
$\overline{q}q$ cannot realize $V_{h=\pm 1}X_{h=\pm 1}$. 
We can easily see that the helicities of $V$ and $X$ do 
not match for $h=\pm 1$ even when $V$ receives $q_{spec}$. 
The only helicity final state compatible with the overall 
angular momentum conservation is still $V_{h=0}X_{h=0}$ 
in the collinear limit. Therefore, the preceding argument 
for the two-body decay $B\to VM$ is carried over to the 
inclusive decay $B\to VX$.  

\noindent
\begin{figure}[ht]
\hskip 5cm
\epsfig{file=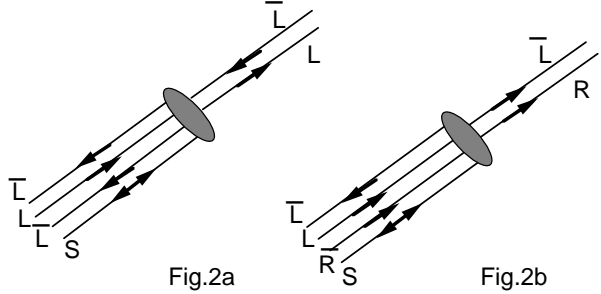,width=8cm,height=4cm} 
\caption{The helicities in the inclusive $B(\overline{b}q)$
decay where an additional hard pair of $\overline{q}q$ is
produced and leads to the final state 
$\overline{q}q\overline{q}q\overline{q}q_{spec}$.
\label{fig:2}}
\end{figure}
         
However, the collinear quark limit becomes a poor approximation  
as $m_X$ increases in the inclusive decay. The transverse quark
momenta ${\bf p}_T$ in $X$ become large with respect to 
${\bf p}_X$ so that the corrections grow with $m_X$. The mass 
correction depends on whether the spectator $q_{spec}$ enters 
$V$ or $X$. For the same reason as in the two-body decay, the 
final helicity state is determined primarily by the meson ($V$) 
or the group of mesons ($X$) that does not receive $q_{spec}$ of 
indefinite helicity. Making an appropriate substitution in the 
mass corrections for the two-body decay, we obtain for 
$m_X\gg m_V$
\begin{eqnarray}
 (1-\Gamma_L)_{mass}\approx\frac{m_V^2 M_B^2}{(M_B^2-m_X^2)^2},
          \;\;(q_{spec}\; {\rm in}\; X),\nonumber \\
 (1-\Gamma_L)_{mass}\approx\frac{m_X^2 m_B^2}{(M_B^2+m_X^2)^2},
          \;\;(q_{spec}\;{\rm in}\; V).   \label{mass}            
\end{eqnarray}
The right-hand sides indicate the orders of magnitude. It is 
difficult even within perturbative QCD to compute their 
coefficients with good accuracy since they depend on the quark 
distributions inside mesons and other details. The coefficients  
are highly dependent on individual decay modes. Nonetheless, 
the rise of $\Gamma_T$ with $m_X^2$, particularly in the case 
that $X$ is produced without the spectator, is an important trend. 
It simply means that the ``small opposite helicity component''
of $O(m_X/E_X)$ ceases to be small when $m_X$ becomes large. 

The orbital motion inside $X$ is not restricted to $l=0$. 
Therefore $l_z$ of $X$ can make up for violation of 
the overall angular momentum conservation when $V$
is formed with $\overline{q_L}q_R$ $(h=+1)$ or 
$\overline{q_R}q_L$ ($h=-1$). In terms of the helicity 
fraction, the $l_z$ correction to $X$ generates the leading
correction that grows rapidly with $m_X$;
\begin{equation}
  (1-\Gamma_L)_{l_z}\approx\frac{m_X^2 m_B^2}{(M_B^2+m_X^2)^2}.
                         \label{lz}
\end{equation}
  
When $\Gamma_T=1-\Gamma_L$ becomes a substantial fraction 
of unity, LDFSI is clearly important.\footnote{
It is possible that $X$ consists of a widely separated pair of
mesons interacting only through SDFSI. In this case, the 
final state is a three-jet state and the decay may be a SD 
process calculable by perturbative QCD for $m_B\to\infty$.
However, such a contribution is suppressed by $O(\alpha_s/\pi)$ 
and not expected to be a significant portion of the inclusive 
decay. One should be able to check by actually examining 
the final states whether it is the case or not.}
As $m_X$ approaches the kinematical upper limit 
corresponding to ${\bf q} = 0$, $\Gamma_L$ should reach 
1/3 according to the limiting behavior of Eq.(\ref{limit}):
\begin{equation}
   \Gamma_L \to 1/3,\;\;{\rm as}\;m_X\to m_{max}.
\end{equation}
Future experiment on the inclusive decay will 
determine $\Gamma_L$ as a function of $m_X$ interpolating 
between $1-O(m_V^2/M_B^2)$ and $\frac{1}{3}$, as sketched 
qualitatively in Fig.3. We should keep in mind that the 
corrections presented here are the expectation based on 
perturbative QCD. It is only a theoretical prediction that 
should be tested by experiment. While the helicity test of 
the charmless decay is of primary interest, no experimental 
data exist on $\Gamma_{T,L}$ for any charmless decay mode
at present.

%\input psfig
%\noindent
\begin{figure}[ht]
\hskip 5cm
\epsfig{file=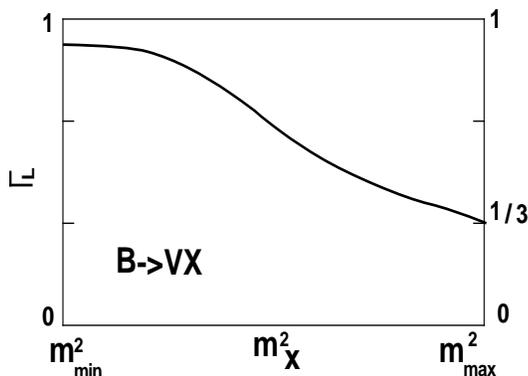,width=7cm,height=5cm} 
\caption{The qualitative behavior of $\Gamma_L$ against $m_X^2$.
While $\Gamma_L=1/3$ at $m_X=m_{max}$ is a kinematical 
constraint, the behavior of $\Gamma_L$ near the small end of 
$m_X$ is only the expectation of perturbative QCD.
\label{fig:3}}
\end{figure}

One problem exists in performing an inclusive measurement of
the charmless decay $B\to VX$. One has to make sure
that $X$ does not contain charm nor hidden charm. Since the
charmless decays are the {\em rare} decays, the region above
the charm threshold for $m_X$ is overwhelmed by the background 
that is much higher in branching. In practice, the charmless 
inclusive decay will be analyzed only in the region 
separated from the charm background by kinematics, that is,
\begin{equation}
              m_X < m_D  \label{range}.
\end{equation}
Above $m_D$, the dominant process is $B\to VX_{\overline{c}}$ 
where $X_{\overline{c}}$ contains an anticharmed meson. 
Fortunately, Eq.(\ref{range}) is the mass range where many
interesting results will be extracted from the charmless 
decay. For the decays into $X_{\overline{c}}$, the helicity 
selection rule holds in a manner almost identical to the
charmless decay. We shall see that the Figure 3 applies to 
$B\to VX_{\overline{c}}$ as well. Therefore, separate tests 
of the rule will be possible with $B\to VX_{\overline{c}}$ 
in the range above $m_X = m_D$. 

As for $V$, reconstruction of $\rho$ from $\pi\pi$ may 
encounter an excessive combinatorial background. If this
happens, $\phi$ will be a clean alternative for $V$ in
the environment of BaBar and Belle.\footnote{
The author owes to R.N. Cahn for this remark.}
As a last resort, we can work on fully reconstructed $B$ 
events with reduced statistics.

\section{Decay into charmed X or charmed V}

We extend the argument for the charmless decay to the 
charmed meson production decay $B\to VX_{\overline{c}}$ 
and $B\to V_{\overline{c}}X$. We ignore here the small 
contribution from the penguin-type processes 
for this class of decays. When $V$ is formed without 
involving the spectator, $V$ carries $h=0$ of 
$\overline{q_L}q_L$ up to the small mass correction given
by the first line of Eq.(\ref{mass}).
The $h=0$ dominance remains true even when an extra 
$\overline{q}q$ pair is produced: Imagine, for instance, 
that $\overline{q_R}$ and $q_R$ are produced secondarily by 
a hard gluon and enter both $V$ and $X$. Then 
$V=\overline{q_L}q_R$ and $X=\overline{c_L}q_L
\overline{q_R}q_{spec}$ can satisfy the overall angular 
momentum conservation only with help of $l_z=+1$ or the 
opposite component of $q_L$ or $\overline{q_R}$. In the case
of $V=\overline{q_R}q_L$ and $X=\overline{c_L}q_R
\overline{q_L}q_{spec}$, both $l_z=-1$ and the opposite 
helicity of $\overline{c_L}$ are needed.\footnote{
The opposite helicity content of $c_L$ is larger; 
$(m_c^2+{\bf p}_T^2)^{1/2}/E_c$ instead of 
$|{\bf p}_T|/E_c$.} 
In the two-body decay where $X_{\overline{c}}$ is 
$\overline{D}^*$ ($l=0$) and $q_{spec}$ enters $D^*$,
therefore, the correction to the $h=0$ rule is dominated
by the mass correction to $V$,  
\begin{equation}
  1-\Gamma_L \approx \frac{m_V^2M_B^2}{(M_B^2-m_X^2)^2}.
          \label{D}
\end{equation} 
This correction will apply to $B^0/\overline{B}^0\to
\rho^{\pm}D^{*\mp}$ since the quark distribution function 
disfavors formation of $\rho^{\pm}$ with the spectator. 
Because of the large branching, experiment already measured 
the helicity fractions with good accuracy for the two-body 
decay $B^0/\overline{B}^0\to\rho^{\pm}D^{*\mp}$ many years 
ago. The experimental result was in agreement with the $h=0$ 
dominance\cite{CLEO};
\begin{equation}
   \Gamma_L  = 0.93\pm 0.05 \pm 0.05. 
\end{equation}
The deviation from unity of $\Gamma_L$ is consistent with
the mass correction ($\approx 0.03$) that we expect from 
Eq.(\ref{D}). Even when $X_{\overline{c}/c}$ is a higher 
state of $l\neq 0$, the correction to the $h=0$ rule is 
determined by $\rho^{\pm}$ and grows rather slowly with $m_X$ 
according to Eq.(\ref{D}) since $q_{spec}$ enters 
$X_{\overline{c}/c}$ in the dominant process of 
$B^0/\overline{B}^0\to\rho^{\pm}X_{\overline{c}/c}$. 

The correction is a little different for the so-called 
color-disfavored decays.  Take $B^0(\overline{b}d)\to
\rho^0\overline{D}^{*0}$ as an example: The $\rho^0$ meson 
must be formed with the spectator when the decay occurs 
through the dominant operator for this decay. The final 
helicity is constrained by $D^{*-}$ and the correction
is $1-\Gamma_L\approx m_X^2M_B^2/(M_B^2+m_X^2)^2$.
Therefore we expect that the correction is larger in 
$B^0\to\rho^0\overline{D}^{*0}$ than in $B^0\to\rho^+D^{*-}$; 
\begin{equation}
   \Gamma_T(B^0\to \rho^0\overline{D}^0) > 
       \Gamma_T(B^0\to\rho^+ D^{*-}). \label{min}
\end{equation} 
The recent measurement\cite{Belle} of the 
factorization-disfavored two-body decays,  
$B^0\to\overline{D}^{(*0)}X^0$ ($X^0=\pi^0, \omega, \eta$)
seems to show that the branching fractions for these decays 
are larger than their lowest-order perturbative QCD 
calculations\cite{Beneke}. The helicity analysis of 
$B^0\to\rho^0 X^0_{\overline{c}}$ and
$K^{*0}X^0_{\overline{c}}$ will help us toward 
better understanding of how much LDFSI is involved here.
  
Let us move to the other inclusive measurement where
a charmed meson is identified instead of a light meson;
$B\to \overline{D}^* X$. There is an experimental
advantage in reconstructing $\overline{D}^*$ through its
soft decay into $\overline{D}\pi$. The $\overline{D}^*$ 
meson can be formed with or without the spectator. With 
the spectator ($\overline{D}^* =\overline{c_L}q_{spec}$), 
the accuracy of the $h=0$ dominance is controlled by the 
helicity of $X$, which is determined by $\overline{q_L}q_L$, 
$\overline{q_L}q_L\overline{q_L}q_L$, 
$\overline{q_L}q_L\overline{q_R}q_R\cdots$. 
The correction is given by the second line of Eq.(\ref{mass}) 
and grows rapidly with $m_X$. On the other hand, when $X$ 
receives the spectator, $X = \overline{q_L}q_{spec}$,
$\overline{q_L}q_{spec}\overline{q}q, \cdots$ can be 
in either $h=+1$ or $0$ with a 50/50 chance. Then it is 
$\overline{D}^* = \overline{c_L}q_L$ that determines the 
final helicity. The dominant helicity is again $h=0$ and 
the correction is given by the first line of Eq.(\ref{mass}),
but the magnitude is large because of the larger opposite
helicity content in $\overline{c_L}$.

Finally we comment on the decays $B\to VX_{\overline{c}c}$
and $V_{\overline{c}c}X$. A pair of $\overline{c}c$ is 
produced by weak interaction and forms one of charmonia 
or turns into $\overline{D}^{(*)}D^{(*)}$. $V$ is most 
likely formed with the spectator since little phase space 
is left for production of a fast pair of 
$\overline{q}q$. In this case, the helicity content is 
determined by $\overline{c_L}c_L$. Since $c_L$ and 
$\overline{c_L}$ are heavy and slow, the opposite helicity 
content of $O(\frac{1}{2}m_{\overline{c}c}/E_{\overline{c}c})$  
does not give an accurate estimate. Nonetheless let us 
stretch for the moment the mass correction formula for 
$\Gamma_T$ such that the coefficient in front be adjusted 
to give the kinematical constraint $\Gamma_T=\frac{2}{3}$ 
at the maximum value of $m_X$. Then the prediction on 
$\Gamma_T$ would be
\begin{equation}
       \Gamma_T \simeq \frac{8}{3} \times 
                  \frac{m_{\overline{c}c}^2 M_B^2}{
                   (M_B^2 + m_{\overline{c}c}^2)^2}, \label{cc}
\end{equation}
where $m_{\overline{c}c}$ is the invariant mass of 
all hadrons but $V$. $X_{\overline{c}c}$ is most likely
one of charmonia. Detailed measurements were made for 
the helicity content of $B\to J/\psi K^*$.  For this decay
mode, Eq.(\ref{cc}) gives a ``correction'' of $\Gamma_L 
\simeq 0.49$. The latest result of the helicity analysis by 
BaBar\cite{BaBar2} can be expressed as,
\begin{equation}
        \Gamma_L = 0.597 \pm 0.028 \pm 0.024, 
\end{equation}
which is not far from 0.49. However, the agreement is probably 
fortuitous since the Lorentz factor $\gamma$ of $J/\psi$ is 
only 1.12 in this decay.

In the decay $B\to J/\psi K^*$, the $K^*$ meson moves with 
$\gamma\simeq 2$. If we make the approximation of $K^*$ being 
fast, $K^*(\overline{s_L}q_{spec})$ can be only in helicity 
$+1$ or 0, not in $-1$.  Therefore $H_{-1}\simeq 0$ is predicted 
for $B\to J/\psi K^*$ if one assumes perturbative QCD for $K^*$. 
The transversity angular analysis\cite{BaBar2} allows two 
solutions, $H_1\gg H_{-1}$ and $H_1\ll H_{-1}$, but cannot 
resolve the twofold ambiguity. At present, experiment
still does not exclude the possibility that 
perturbative QCD is applicable to the light meson side 
($K^*$) of the decay.\footnote{
Following earlier experimental papers\cite{others}, the BaBar 
analysis\cite{BaBar2} quotes only one solution, 
$\phi_{\parallel}-\phi_{\perp}\simeq \pi$, which would 
lead to $H_1\ll H_{-1}$ in the ordinary sign convention 
chosen in its reference\cite{Dighe}. It might look
as if the BaBar result were in direct conflict with 
the prediction of perturbative QCD for $K^*$. In fact, 
the other solution $\phi_{\parallel}-\phi_{\perp}\simeq 0$ 
leading to $H_1\gg H_{-1}$ is also allowed by this experiment,
though not explicitly quoted as such\cite{Suzuki2}.
Therefore no conclusion can be drawn from this experiment 
as to which is larger between $H_1$ and $H_{-1}$  
in $B\to J/\psi K^*$. The same comment applies to the latest 
Belle analysis\cite{Belle2}.} 

The Belle Collaboration recently measured the branching fraction 
for the factorization-suppressed decay $B\to\chi_0 K$\cite{Belle} 
at the level comparable with the factorization-favored decays 
$B\to\eta_c K$, $J/\psi K$, and $\chi_1 K$. It shows that the 
simple factorization clearly fails in the decay $B\to$ charmonium.

The decay $B\to\overline{D}^*D^*$ is being analyzed at 
the B-factories. The branching fraction was reported for
$D^{*+}D^{*-}$\cite{BaBar3}. After accumulation of more events,
the helicity analysis will become feasible. Comparison of this
decay with $B\to J/\psi K^*$ may provide an additional useful 
information about dynamics in $b\to c\overline{c}q$. 

\section{Higher spin ($J\geq 2$)}

  The helicity test can be performed for higher-spin inclusive 
processes $B\to MX \to abX$ with $J\geq 2$ for $M$. For $J=0$ 
for $a$ and $b$, the differential decay rate in the $B$ rest 
frame takes the form,
\begin{equation}
   \frac{d\Gamma}{dq_0 d\cos\theta} \propto |{\bf q}|
            \sum_{\lambda=-J}^J H_{\lambda}(m_X^2)
            |d^J_{\lambda,0}(\theta)|^2,
\end{equation}
where $\lambda$ is the helicity of $M$.
The momentum ${\bf q}$ and the angle $\theta$ 
are defined in the same way as in Eq.(\ref{angular}).
In the case of $J\neq 0$ for $a$ and/or $b$, an additional 
$\lambda$ dependence enters through the decay $M\to a + b$.  
The dominant helicity structure function is $H_0$, then
$H_{\pm 1}$ for both $B$ and $\overline{B}$ decays, since the 
$l_z$ correction to $M$ contributes to $H_1$ and $H_{-1}$ 
in the same order. If perturbative QCD is valid, the function 
$H_{\lambda}$ with $|\lambda|\geq 2$ cannot arise without the
$l_z$ correction.  $H_h$ with $|h|\geq 2$ beyond the $l_z$
correction will be a clear evidence for LDFSI. As $m_X$ tends 
to its maximum value, $\Gamma_L$ should approach $1/(2J+1)$.
In the decay $B\to f_2 X \to \pi\pi X$, for instance, 
the angular dependence $|d^2_{\pm 2 0}|^2 = 
\frac{3}{8}(1-\cos^2\theta)^2$ appears as $f_2$ slows down.
Appearance of $(1-\cos^2\theta)^2$ indicates that the orbital 
angular momentum of $\overline{q}q$ inside $f_2$ becomes 
important in the $B$ rest frame. One might think of attributing 
appearance of $|h|\geq 2$ to possible breakdown of the 
$\overline{q}q$ description of $f_2$. But it is unlikely 
in the face of the static quark model: the $\overline{q}q$ 
description of low-lying mesons works well both in the infinite 
momentum limit and in the static limit albeit the physical 
nature of quarks is different between the two limits.  As 
${\bf q}\to 0$, all $l_z$ states of $f_2$ are equally produced 
and $\Gamma_L$ should approach 1/5.

\section{Comparison with other tests}

Various tests have so far been proposed concerning validity of 
the factorization. The most straightforward is to compute 
as many decay amplitudes as possible with theoretical resources 
at hand. In some simple cases we are fortunate to have only a 
single dominant decay process in the factorization limit. 
An example is $B^0\to D^-\pi^+$. Otherwise a decay amplitude for 
a given process is sum of competing contributions of more than 
one decay process. Once short-distance QCD corrections are 
included, the quark operators producing mesons are nonlocal. 
Then we need to know not only the decay constants, the wave 
functions at origin, but also the entire light-cone quark 
distribution functions in order to obtain a single decay 
amplitude. Furthermore, the relevant energy scale of the QCD 
coupling $\alpha_s(E)$ can take different values depending on 
how and where it appears. Therefore a final number for a total 
decay amplitude is sensitive to small theoretical uncertainties 
of each contribution particularly when different terms enter 
with different signs. These added uncertainties make comparison 
of theory with experiment less decisive. For this reason we 
give up here attempting numerical estimate of the coefficients 
of the corrections to the $h=0$ dominance rule even for the 
simplest two-body decay $B\to 1^-1^-$. 

  A while ago Ligeti {\em et al.}\cite{Ligeti} proposed 
a test of the factorization in the decay $B\to
\overline{D}^{(*)}X$. They proposed to compare the $m_X$ 
distribution of this inclusive decay with the $m_{l\nu}$ 
distribution of the semileptonic decay $B\to 
\overline{D}^{(*)}\overline{l}\nu$. It appears to be 
a clean test.  In order for this test to work, however,
$X$ must be produced from a single weak current just 
as $\overline{l}\nu$ is. Therefore, it applies to 
$B^0\to D^{(*)-}X^+$ (and the conjugate) through the 
dominant decay operator, but not to $B^+\to D^{(*)0}X^+$ 
(and the conjugate) since $X^+$ can pick either the current 
quark $u$ or the spectator $u$ in the $B^+$ decay. Only 
the neutral $B$ decay is possibly related to the 
semileptonic decay. The most important difference from
our test is that the comparison with the semileptonic
decay tests only validity of the factorization 
before the perturbative QCD improvement. The SDFSI 
surely plays a significant role in the final state and
breaks down the similarity between the nonleptonic and the
semileptonic decay.  An alternative to this test was 
proposed for two-body decays and importance of spin was 
mentioned\cite{Hiller}, but it is not free of the 
uncertainties and complications in theoretical computation. 
In contrast, the inclusive helicity measurement tests not 
just the lowest-order factorization but its perturbative 
QCD corrections to all orders independent of theoretical 
details. It will provide us with  
an important information as to how much long-distance QCD 
interactions enter a given process and allow us to use 
it for related processes. A negative side of the helicity 
test is, of course, the common drawback of LDFSI that 
after LDFSI is found, we cannot compute phases nor magnitudes 
of decay amplitudes from the first principle. However, 
just measuring CP violations beyond the $B^0$-$\overline{B}^0$ 
mixing effect will be important even if we cannot easily relate 
it to fundamental parameters of theory.  Only when LDFSI is 
significant, do we have a chance to detect a direct CP 
violation from particle-antiparticle asymmetry. The helicity 
test will hopefully tell us which decay modes we should  
go after for search of direct CP violations. 

\section{Acknowledgment}
This work was supported in part by the Director, Office of 
Science, Office of High Energy and Nuclear Physics, Division 
of High Energy Physics, of the U.S. Department of Energy under 
contract DE-AC03-76SF00098 and in part by the National Science 
Foundation under grant PHY-0098840.

%%\input psfig
%\noindent
%\begin{figure}
%%\epsfig{file=fig01.eps,width=0.47\textwidth}
%\epsfig{file=incf1.eps,width=8cm,height=4cm} %figurename =incf1
%\caption{The quark helicities in the two-body $B(\overline{b}q)$ 
%decay. $L$ ($\overline{L}$) and $R$ ($\overline{R}$) denote left 
%and right chiral quarks (antiquarks), respectively. The spectator 
%quark $q_{spec}$ is denoted by $S$. The thick arrows represent 
%the dominant spin directions. (a) The case of $\overline{b}\to
%\overline{q_L}q_L\overline{q_L}$. (b) The case of $\overline{B}
%\to\overline{q_L}q_R\overline{q_R}$.
%\label{fig:1}}
%\end{figure}

%%\input psfig
%\noindent
%\begin{figure}
%%\epsfig{file=fig01.eps,width=0.47\textwidth}
%\epsfig{file=incf2.eps,width=8cm,height=5cm} % figurename =incf2
%\caption{The helicities in the inclusive $B(\overline{b}q)$ 
%decay where an additional hard pair of $\overline{q}q$ is 
%produced and leads to the the $\overline{q}q\overline{q}q
%\overline{q}q_{spec}$ final state.
%\label{fig:2}}
%\end{figure}

%%\input psfig
%\noindent
%\begin{figure}
%%\epsfig{file=fig01.eps,width=0.47\textwidth}
%\epsfig{file=incf3.eps,width=7cm,height=5cm} %figurename =incf3
%\caption{The qualitative behavior of $\Gamma_L$ against $m_X^2$.
%While $\Gamma_L=1/3$ is a kinematical constraint, the behavior
%of $\Gamma_L$ near the small end of $m_X^2$ is the expectation 
%of perturbative QCD.
%\label{fig:3}}
%\end{figure}

\end{document}